\documentclass[aps,prl,twocolumn,superscriptaddress]{revtex4}

\usepackage{mathrsfs}
\usepackage{graphicx}
\usepackage{bm}
\usepackage{threeparttable}
\usepackage{amsmath,amssymb}
\usepackage{color}
\usepackage{epstopdf}
\usepackage[hidelinks]{hyperref}
\usepackage{changes}
\colorlet{Changes@Color}{red}

\begin{document}

\title{Revisiting fluid-wall interfacial tension}

\author{Longfei Li}
\affiliation{Beijing National Laboratory for Condensed Matter Physics and Laboratory of Soft Matter Physics, Institute of Physics, Chinese Academy of Sciences, Beijing 100190, China}

\author{Mingcheng Yang}
\email{mcyang@iphy.ac.cn}
\affiliation{Beijing National Laboratory for Condensed Matter Physics and Laboratory of Soft Matter Physics, Institute of Physics, Chinese Academy of Sciences, Beijing 100190, China}
\affiliation{School of Physical Sciences, University of Chinese Academy of Sciences, Beijing 100049, China}
\affiliation{Songshan Lake Materials Laboratory, Dongguan, Guangdong 523808, China}

\begin{abstract}
A fluid in contact with a flat structureless wall constitutes the simplest interface system, but the fluid-wall interfacial tension cannot be trivially and even unequivocally determined due to the ambiguity in identifying the precise location of fluid-wall dividing surface. To resolve this long-standing problem, we here derive the interfacial tension from two independent routes without needing the identification of dividing surface. The first one exploits a natural idea that the interfacial profiles of intensive quantities should remain perfectly invariant when deforming the fluid-wall system just to change its interface area. The second one considers the fluid-wall system as the limit of a fluid under a finite external potential field. By calculating the work required to create a differential interface area, the two methods yield exactly the same interfacial tension. Thus, our work provides strong evidence that the fluid-wall interfacial tension can be unambiguously quantified.
\end{abstract}

\pacs {}
\maketitle

\section{Introduction}

The fluid-solid interfacial tension plays a pivotal role in plentiful phenomena occurring on solid surface such as wetting, spreading, adhesion and capillary action~\cite{Gennes1985rmp,Bonn2009rmp,Starov2019book,Lee2013book,Gennes2004book}, and is vital to numerous applications. However, its direct quantification remains challenging because of the non-rearrangement of solid particles and the anisotropy of solid structure~\cite{Shuttleworth1950book,Defay1966book,Navascues1979rpp}. To simplify the study of the real fluid-solid interfacial tension, the solid phase is often approximately treated as a flat structureless wall~\cite{Navascues1977mp,Navascues1979rpp,Jones1999fd,Leroy2015lang,Fan2020prl}. Even for such a minimal fluid-wall interface, the interfacial tension cannot be unequivocally determined due to the ambiguity in identifying the precise location of the dividing surface between the fluid and wall phases, which is related to the volume of a confined fluid~\cite{Roth2010jpcm,Reindl2015pre,Martin2018jcp,Davidchack2014mp,Dong2021pnas}.

Up to now, there are mainly two ways to directly calculate the fluid-wall interfacial tension $\gamma_w$. One is based on its mechanical meaning~\cite{Metiu1971A,Navascues1977mp,Navascues1979rpp,Adams1991mp,Jones1999fd,
Miguel2006mp,Benjamin2012jcp,Miguez2012jcp,Lukyanov2013lang,Leroy2015lang,
Shou2016jcp,Wandelt2016book,Fan2020prl,Dong2021pnas,Dong2023nc,Nijmeijer1990pra,Stillinger1962jcp}, and the resulting interfacial tension can be expressed in the unified form,
\begin{equation}
	\label {eq1}
	\begin{split}
	\gamma_w \! = \! \int_{-\infty}^{x_0} [P_{\alpha} - P^T(x)] {\rm d}x + \! \int_{x_0}^{+\infty} [P_{\beta} - P^T(x)] {\rm d}x.
	\end{split}
\end{equation}
Here, the $x$ axis is perpendicular to the interface layer and $x_0$ is called the \textit{Gibbs dividing surface} (GDS)~\cite{Gibbs1928book} that is an imaginary mathematical surface translating the actual continuous interfacial region into two contacting homogeneous bulk phases (i.e., $\alpha$ and $\beta$). For the present fluid-wall situation, $P_{\alpha}$ and $P_{\beta}$ respectively denote the pressures inside the wall phase and in the bulk-phase fluid ($P_{\alpha} = 0$ due to the vanishing fluid density inside the wall), and $P^T(x)$ is the actual local tangential pressure of the fluid. Equation~(\ref{eq1}), which is originally obtained for planar gas-liquid interface systems, can be easily derived by calculating the work required to create a differential interface area, once the GDS is imposed.

\begin{figure}[b]
	\includegraphics[width=0.4\textwidth]{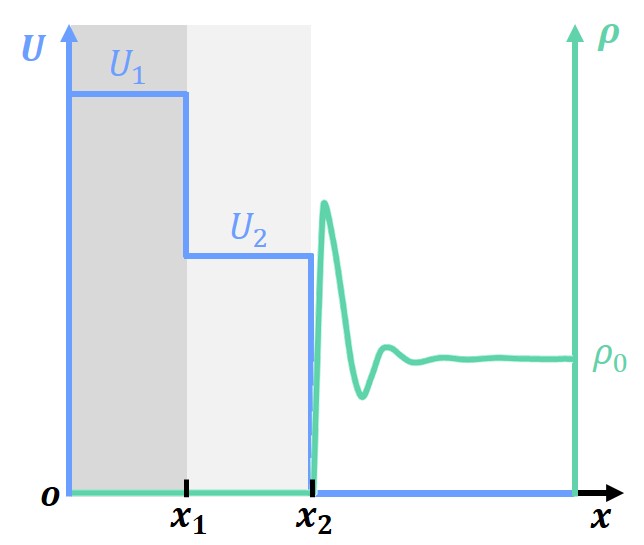}
	\caption{Sketch of a fluid-wall system with a hypothetical piecewise-type wall potential $U(x)$ (blue curve). For clarity, the corresponding schematic density profile of fluid particles (green curve) is also depicted.}
	\label{Fig1}
\end{figure}

The use of GDS is a standard way to investigate the interfacial tension and largely simplifies the treatment for real interface systems. Nevertheless, the position of GDS is not unique, and any
mathematical surface parallel to the actual interface layer can be chosen to be the GDS~\cite{Rowlinson2013book}. It is well known that the precise position of GDS is irrelevant to the interfacial tension of a gas-liquid system with a planar interface (where the coexisting gas and liquid have the same pressure). However, for the fluid-wall system, since $P_{\alpha} \ne P_{\beta}$, $\gamma_w$ clearly depends on the location of GDS [see Eq.~(\ref{eq1})]. A fundamental problem in the determination of $\gamma_w$ is thus to properly impose the GDS. A seemingly reasonable choice is to set GDS ($x_0$) at the location where the wall potential just right reaches infinity, as done in almost all previous studies~\cite{Metiu1971A,
Navascues1977mp,Navascues1979rpp,Adams1991mp,Jones1999fd,Miguel2006mp,Benjamin2012jcp,Miguez2012jcp,Lukyanov2013lang,Leroy2015lang,
Shou2016jcp,Wandelt2016book,Fan2020prl,Dong2021pnas,Dong2023nc,Nijmeijer1990pra}, leading to $\gamma_w \! = \! \int_{x_0}^{+\infty} [P_{\beta} - P^T(x)] {\rm d}x$ or its equivalent. Although the GDS thus chosen coincides with the maximum range that can be explored by the fluid particles, the resulting $\gamma_w$ still is indeterminate, as will be demonstrated as follows.

We consider a planar fluid-wall interface, where the wall interacts with the fluid particles via a hypothetical potential $U(x)$, as sketched by the blue curve in Fig.~\ref{Fig1}. The $U(x)$ is a piecewise function: $U(x) = U_1$ for $x \le x_1$, $U(x) = U_2$ for $x_1 < x \le x_2$, and $U(x)=0$ otherwise. Further, we take $U_1$ as infinity and $U_2$ as a large finite value (say, $U_2 = 10^4 k_{\tt B}T$ with $k_{\tt B}T$ the thermal energy). In this case, the fluid particles are forbidden to enter the region $x<x_2$, as sketched by the fluid density profile (the green curve in Fig.~\ref{Fig1}). Consequently, the effects of the wall potentials with different $x_1$ ($0< x_1 < x_2$) are actually identical, namely, the physical properties of the interfacial system do not depend on $x_1$. However, if $x_1$ is chosen as the GDS, then the interfacial tension will obviously change with $x_1$ [see Eq.~(\ref{eq1})]. This situation is self-contradictory, implying that the GDS should not be taken where the wall potential just reaches infinity. Moreover, the GDS cannot be placed at $x_2$ either, otherwise this choice will raise an unsolvable problem: how large does $U_2=U(x_2)$ need to be at least? 

Unlike Eq.~(\ref{eq1}), some studies regard the integration of the difference between local normal and tangential pressures as the interfacial tension, $\gamma_w=\int_{-\infty}^{\infty} [P^N(x) - P^T(x)] {\rm d}x$~\cite{Gray2011book,Henderson1983mp,
Henderson1984mp,Henderson1985mp,Henderson1992book,Swol1986fs,Thompson1993jast}. Although this expression is independent of the GDS, it does not correspond to the work required to create a differential interface area, i.e., the mechanical definition of $\gamma_w$.

The other way to directly calculate interfacial tension is to exploit thermodynamic potentials and their relations~\cite{Reindl2015pre,Roth2010jpcm,
Schimmele2007jcp,Bellemans1962p,Mcquarrie1987mp,Bryk2003pre,
Konig2004prl,Nijmeijer1990jpa,Yang2013jcp,Yang2015mp,Neumann2010book,Blokhuis2007jcp,Urrutia2016jcp,Sullivan1981jcp,Bragado1984pra,Navascues1979mp,
Saville1977ft,Heni1999pre,Martin2018jcp,Martin2020jpcb,Dash2012book,
Benjamin2013jcp,Davidchack2018jcp,Kern2014jcp,Laird2010jcp,Reiss1959jcp,Reiss1965book}. The surface excess grand potential is widely employed to quantify $\gamma_w$, that is the system grand potential subtracting the bulk contribution~\cite{Reindl2015pre,
Roth2010jpcm,Schimmele2007jcp,Bellemans1962p,Mcquarrie1987mp,
Bryk2003pre,Konig2004prl,Nijmeijer1990jpa,Yang2013jcp,Yang2015mp,
Neumann2010book,Blokhuis2007jcp,Sullivan1981jcp,Urrutia2016jcp}. The derivative of Helmholtz free energy with respect to interfacial area, with fixed fluid volume, particle number and temperature, can also give $\gamma_w$~\cite{Bragado1984pra,Navascues1979mp,Saville1977ft,Heni1999pre}. In addition, the Gibbs-Cahn thermodynamic integration can be used to determine $\gamma_w$~\cite{Martin2018jcp,Martin2020jpcb,Dash2012book,
Benjamin2013jcp,Davidchack2018jcp,Kern2014jcp,Laird2010jcp}. All these schemes require specifying the fluid volume without exception. However, unfortunately, it is difficult to unambiguously identify the volume or border (i.e., the dividing surface) of a confined fluid~\cite{Yang2013jcp,Yang2015mp,Nijmeijer1990jpa,Martin2018jcp}. An intuitive and widely-adopted choice, that takes the fluid border at the location of infinite wall potential, encounters the same problem as the aforementioned GDS. Even for a purely hard wall, it is unclear whether the fluid volume should be measured from the wall surface or from the center of the fluid particle in contact with the wall~\cite{Roth2010jpcm,Martin2018jcp,Reindl2015pre,Davidchack2014mp}, and different prescriptions may give rise to  opposite-sign $\gamma_w$~\cite{Roth2010jpcm,Reindl2015pre,Heni1999pre}.

From the above discussions, it is still an open fundamental question to unambiguously determine the interfacial tension of a fluid-wall system. To tackle this long-standing difficulty, we here revisit the fluid-wall interfacial tension through two independent routes that do not need the identification of the GDS and thus avoid the related ambiguity. First, we heuristically derive the formula of $\gamma_w$ from its mechanical definition, by requiring that the interfacial profile of any intensive quantity perfectly remains invariant during creating an interface area. This very natural requirement, which is ignored in previous studies, has to be satisfied to perform a proper derivation. Then, we obtain the same $\gamma_w$ rigorously, by considering the fact that the fluid-wall system can be treated as the limit of a fluid in a finite external potential field, where the fluid volume is well-defined.

\begin{figure}[b]
	\includegraphics[width=0.47\textwidth]{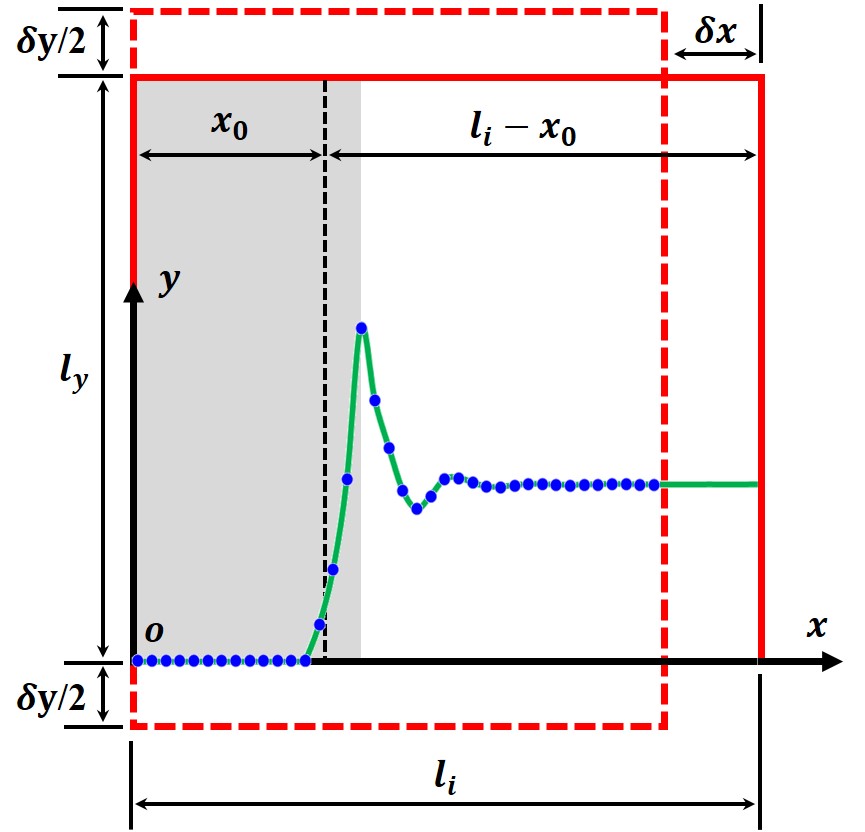}
	\caption{Schematic of the fluid-wall interface region used to derive $\gamma_w$ in route I. The red solid and dashed squares separately denote the original and deformed interface regions with the $x$ axis perpendicular to the wall, the green solid and blue dotted curves are the corresponding density profiles $\rho(x)$, and the shaded region represents the wall region. Here, the vertical black dashed line, $x = x_0$, corresponds to the inferred fluid-wall GDS.}
	\label{Fig2}
\end{figure}

\section{Theoretical route I}

For convenience, the following derivation is restricted to two dimensions, and generalizing our results to three-dimensional systems is straightforward. Specifically, we consider a fluid confined by two vertical planar structureless walls. The system can be divided into a homogeneous bulk fluid phase and two inhomogeneous interface regions. The density and pressure of the bulk fluid are $\rho_0$ and $P_0$, respectively. To proceed, we focus on the interface region near the left wall, with a length of $l_i$ and a width of $l_y$, as sketched in Fig.~\ref{Fig2}. Notably, the left and right boundaries of the interface region are taken deep into the wall and the bulk fluid, respectively. So, the interfacial profile of any intensive quantity will saturate to the corresponding bulk-phase values at the positions far from the boundaries of the interface region. For example, the density $\rho(x)$ separately reaches zero or $\rho_0$ before crossing the left or right boundary of the interface region, as plotted by the green solid curve in Fig.~\ref{Fig2}. 

To derive the formula of $\gamma_w$ from the work related to the increase of interface area, we isothermally and quasistatically extend the interface region by a differential length $\delta y$ in the $y$ direction while compress its right boundary by $\delta x$ in the $x$ direction, as displayed by red dashed square in Fig.~\ref{Fig2}. To determine the relation between $\delta x$ and $\delta y$, we notice that any intensive quantity profile (e.g., density profile, as represented by the blue dotted curve in Fig.~\ref{Fig2}) of the deformed interface region should keep the same as the counterpart of the original interface. This condition is vitally important, since, otherwise, the system intensive quantities (hence the interfacial and bulk properties) will change. In other words, only in this situation can it be ensured that we are examing the originally specified interface. As a result of the invariant density profile, we have
\begin{equation}
	\label {eq2}
	\begin{split}
		l_y \int_0^{l_i} \! \rho(x) \, {\rm d}x  = (l_y + \delta y) \int_{0}^{l_i- \delta x} \! \rho(x) \, {\rm d}x,
	\end{split}
\end{equation}
in which the conservation of particles has been employed. The work $\delta W$ done on the interface region in this process readily reads
\begin{equation}
	\label {eq3}
	\begin{split}
		\delta W = P_0 l_y \delta x - \delta y\int_0^{l_i} P^T(x) \, {\rm d}x,
	\end{split}
\end{equation}
with the local tangential pressure $P^T(x)$ being quantified in the $y$ direction.

Generally, the work $\delta W$ done during the deformation of an interface system is divided into two parts, $\delta W= \gamma_w \delta A - P \delta V$. Here, the first term, $\gamma_w \delta A$, corresponds to the work done by the interfacial tension, and the second term, $-P \delta V$, is associated with the change in the fluid volume. Now, a crucial question is what the value of $\delta V$ is under the deformation that does not change the density profile. Unfortunately, for the fluid-wall system, to obtain $\delta V$ needs to explicitly identify the volume of the confined fluid, which inevitably faces the ambiguity as stated in the \emph{Introduction}. Thus, it seems that $\gamma_w$ could not be determined unambiguously. Nevertheless, we notice that the condition Eq.~(\ref{eq2}), which does not involve the fluid volume, is sufficient to describe the deformation of a fluid-wall system. Such a deformation is the \textit{minimal} one that solely creates a differential fluid-wall interface area without other effect. It is very instructive to compare this minimal deformation of the fluid-wall system to that of a gas-liquid interface system, whose volume is well-defined. For a gas-liquid interface, the minimal deformation (only changing the interface area) additionally needs to keep the volume unchanged, besides the invariant density profile.

While the volume of the confined fluid cannot be unequivocally identified \textit{a priori}, the above observations strongly suggest that the  \textit{minimal} deformation [Eq.~(\ref{eq2})] of the fluid-wall interface natively conserves the fluid volume, i.e., $\delta V=0$. Following this argument, plugging Eq.~(\ref{eq2}) into Eq.~(\ref{eq3}) together with $\gamma_w = \delta W / \delta A$ ($\delta A$ is replaced by $\delta y$ in the present 2D system), the interfacial tension reads
\begin{equation}
	\label {eq4}
	\begin{split}
		\gamma_w = \int_0^{l_i} \left[P_0 \frac{\rho(x)}{\rho_0} - P^T(x) \right]  {\rm d}x.
	\end{split}
\end{equation}
Equation~(\ref{eq4}) is our central result, and its validity will be confirmed by a rigorous derivation below. Moreover, because $\gamma_w$ in Eq.~(\ref{eq4}) is obtained from a purely mechanical way, it is valid in both equilibrium and out-of-equilibrium systems (including active fluids). When using Eq.~(\ref{eq4}) to calculate the interfacial tension of the system in Fig.~\ref{Fig1}, $\gamma_w$ does not hinge on the specific value of $x_1$ any more (given that $0< x_1 < x_2$) and hence is uniquely determined.

Although we do not prescribe the fluid-wall GDS  in the above calculation, a reasonable GDS, hence the fluid volume in the interface region, can be inferred from $\delta V=0$ during the natural minimal deformation. The conservation of the fluid volume before and after the deformation means
\begin{equation}
	\label {eq5}
	\begin{split}
		(l_i - x_0) l_y = (l_i - x_0 - \delta x) (l_y + \delta y),
	\end{split}
\end{equation}
with $x_0$ denoting the location of the GDS (the fluid-wall border) represented by the black dashed line in Fig.~\ref{Fig2}. Note that the effective length of the fluid region is $ l_i - x_0$, instead of $l_i$. Combining Eq.~(\ref{eq5}) and Eq.~(\ref{eq2}) results in
\begin{equation}
	\label {eq6}
	\begin{split}
		\int_0^{x_0} \rho(x) {\rm d}x = \int_{x_0}^{l_i} [\rho_0 - \rho(x)] {\rm d}x.
	\end{split}
\end{equation}
Clearly, the fluid border $x_0$ obeying Eq.~(\ref{eq6}) corresponds to the so-called \textit{equimolar dividing surface}~\cite{Ono1960book,Rowlinson2013book}. This result means if using Eq.~(\ref{eq1}) to calculate $\gamma_w$ then the GDS should be taken as the equimolar dividing surface, unlike the vast majority of literature in which the fluid border is placed where the wall potential just reaches infinity. Indeed, with the equimolar dividing surface, Eq.~(\ref{eq1}) becomes Eq.~(\ref{eq4}). To our best knowledge, in an early work~\cite{Stillinger1962jcp}, Stillinger and Buff chose (without justification) the equimolar dividing surface as the fluid-wall GDS, as they found that such a choice allows a simpler description of the density of the hypothetical uniform fluid phase. In the next section, we will derive the interfacial tension in Eq.~(\ref{eq4}) rigorously.

\section{Theoretical route II}

\begin{figure}[t]
	\includegraphics[width=0.4\textwidth]{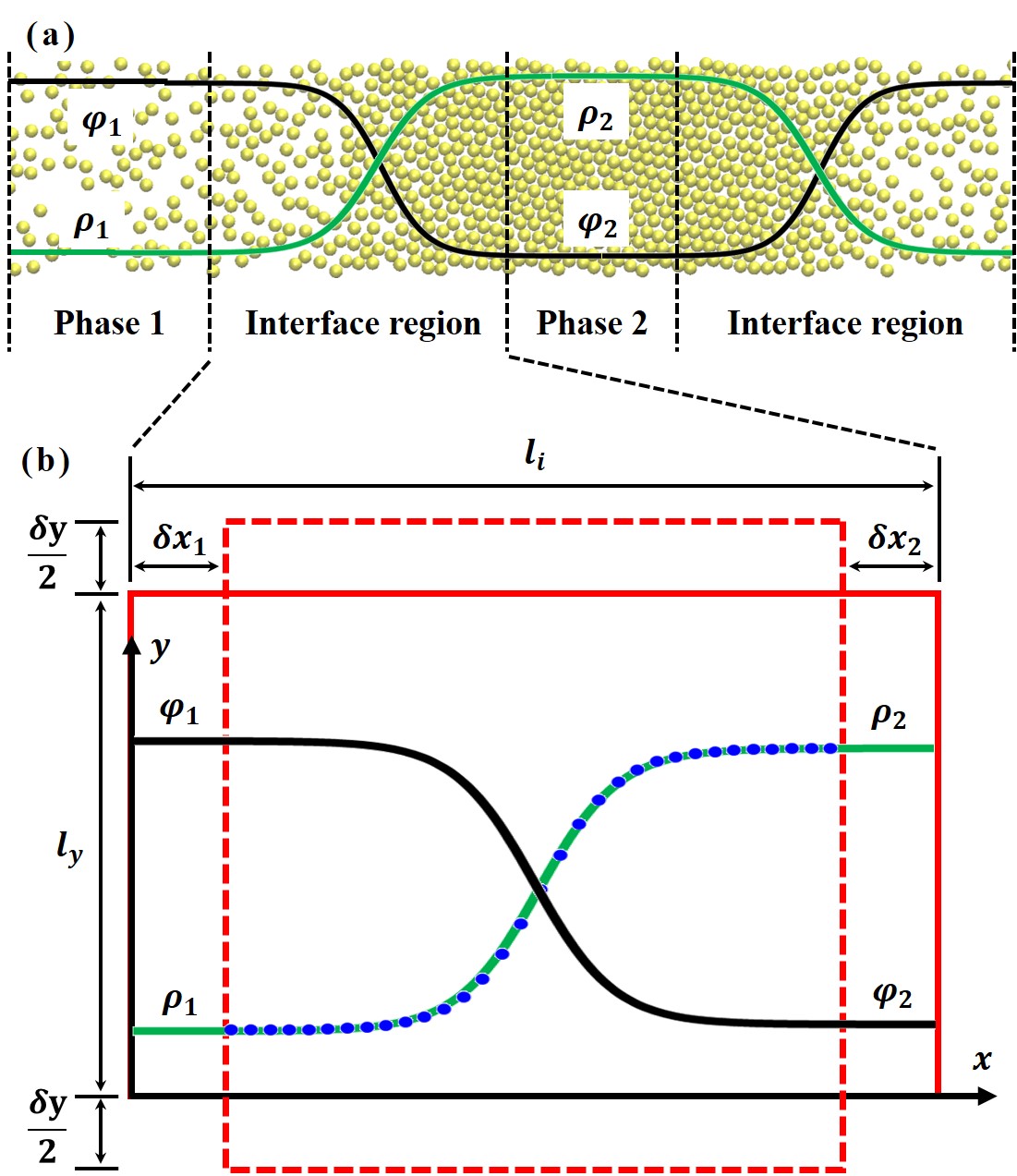}
	\caption{(a) The `coexisting' dilute and dense fluids induced by a piecewise-type external potential $\varphi(x)$ (black solid curve), where the interface regions are sandwiched between the dilute and dense phases. (b) Schematic of the deformed interface region used to derive the interfacial tension in route II. Here, the red solid and dashed squares, respectively, represent the original and deformed interface regions, and the green solid and blue dotted curves are the corresponding density profiles $\rho(x)$.}
	\label{Fig3}
\end{figure}

We consider a periodic fluid system under a piecewise-type external potential field $\varphi(x)$ with the maximum value $\varphi_1$ and minimum value $\varphi_2$, as sketched in Fig.~\ref{Fig3}(a). In this case, a homogeneous dilute fluid of density $\rho_1$ `coexists' with  a homogeneous dense fluid of density $\rho_2$, sandwiching a nonuniform interface region with a density profile $\rho(x)$. The bulk pressures of the dilute and dense phases are $P_1$ and $P_2$, respectively. Similar to route I, we concentrate on one interface region of the whole system, as sketched in Fig.~\ref{Fig3}(b) by the red solid square of length $l_i$ and width $l_y$. Also, the boundaries of the interface region are taken deep into each bulk phase, so that the density profile (the green solid curve in Fig.~\ref{Fig3}(b)) saturates to the corresponding bulk-phase values before crossing the boundaries of the interface region.

To create a differential interface area, we isothermally and quasistatically extend the interface region in the $y$ direction by $\delta y$, meanwhile compress inwards the left and right boundaries of the interface region in the $x$ direction by $\delta x_1$ and $\delta x_2$, respectively, as depicted by the red dashed square in Fig.~\ref{Fig3}(b). In the same spirit as route I, the interfacial deformation should meet the requirement of invariant density profile. For simplicity, the external potential field in the interface region keeps fixed, such that the resulting density profile [the blue dotted curve in Fig.~\ref{Fig3}(b)] overlaps with the original one. Thus, with the particle conservation, we have
\begin{equation}
	\label {eq8}
	\begin{split}
		l_y \int_0^{l_i} \! \rho(x) \, {\rm d}x  = (l_y + \delta y) \int_{\delta x_1}^{l_i- \delta x_2} \! \rho(x) \, {\rm d}x.
	\end{split}
\end{equation}
Moreover, the fluid volume of the present system can be unambiguously determined. In order to exclude the work related to the volume change, we require the deformation to conserve the volume, thus yielding
\begin{equation}
	\label {eq7}
	\begin{split}
		l_i l_y = (l_i - \delta x_1 - \delta x_2)(l_y + \delta y).
	\end{split}
\end{equation}
The work done on the interface region, which exclusively arises from the interfacial tension, reads
\begin{equation}
	\label {eq9}
	\begin{split}
		\delta W = P_1 l_y \delta x_1 + P_2 l_y \delta x_2 - \delta y\int_0^{l_i} P^T(x) \, {\rm d}x,
	\end{split}
\end{equation}
Substituting Eqs.~(\ref{eq8}) and (\ref{eq7}) into (\ref{eq9}), and using the mechanical definition of interfacial tension, one can readily obtain
\begin{equation}
	\label {eq10}
	\begin{split}
		\gamma = \int_0^{l_i} \! \left[ P_1\frac{\rho_2 - \rho(x)}{\rho_2 - \rho_1} + P_2 \frac{\rho(x) - \rho_1}{\rho_2 - \rho_1} - P^T(x) \right] {\rm d}x.
	\end{split}
\end{equation}
Equation (\ref{eq10}) is very general and valid for any piecewise-type external field. When the maximum and minimum values of the external potential are separately taken as infinity and zero ($\varphi_1=\infty$ and $\varphi_2=0$), both $P_1$ and $\rho_1$ vanish, and the present system exactly becomes a fluid-wall interface. In this situation, Eq.~(\ref{eq10}) trivially reduces to Eq.~(\ref{eq4}). Here, Eq.~(\ref{eq10}), hence Eq.~(\ref{eq4}), is obtained without any ambiguity and approximation, therefore the fluid-wall interfacial tension can be unequivocally quantified. In the above calculation, for convenience, the  external potential field is considered to be fixed. If the external field wholly experiences a displacement along the $x$ direction, it will do extra work, by including which Eq.~(\ref{eq10}) can also be derived in a similar manner.

As a case study, we use Eq.~(\ref{eq4}) to calculate $\gamma_w$ for the simplest interface system, an ideal gas in contact with a flat wall. With the bulk-gas pressure $P_0 = k_{\rm B}T \rho_0$ and the local tangential pressure $P^T(x)=k_{\rm B}T \rho(x)$, Eq.~(\ref{eq4}) predicts that the interfacial tension between the ideal gas and wall is equal to zero, independent of the specific form of gas-wall interaction. This null $\gamma_w$ is consistent with the fact that the interfacial tension microscopically is attributed to the interactions between fluid particles in a nonuniform and anisotropic environment. Although $\gamma_w=0$ for the ideal gas-wall interface is a trivial result, it provides a preliminary criteria to roughly assess the appropriateness of the formula of the fluid-wall interfacial tension.

\section{Conclusion}

Based on two independent routes without needing to identify the location of fluid-wall dividing surface, we derive the mechanical expression for the fluid-wall interfacial tension. The obtained expression applies equally well to both equilibrium and nonequilibrium systems. In particular, for the extensively studied active fluids in confined environments~\cite{Zakine2020prl,Turci2021prl}, the active fluid-wall interfacial tension can be directly calculated from Eq.~(\ref{eq4}) by using the intrinsic pressure of active fluid, that excludes the swimming pressure. Our results indicate that the fluid-wall interfacial tension can be uniquely determined, thus solving the long-standing related ambiguity.

\section{Acknowledgment}
We thank Rudolf Podgornik, Wei Dong, Yongfeng Zhao and Zihao Sun for helpful discussions. This work was supported by the National Natural Science Foundation of China (No. 12274448, T2325027).

\bibliographystyle{apsrev}
\bibliography{manuscript}


\end{document}